\documentclass[a4paper,twocolumn]{esapub} 

\usepackage{natbib}
\usepackage{graphicx}

\title{Variability of X-ray pulsars in a hard energy band observed 
with \textit{INTEGRAL}}
\author[1]{A.Lutovinov}
\author[1]{S.Tsygankov}
\author[1,2]{M.Revnivtsev}
\author[3]{M.Chernyakova}
\author[4,5]{I.Bikmaev}
\author[1]{S.Molkov}
\author[1]{R.Burenin}
\author[1]{M.Pavlinsky}
\author[4,5]{N.Sakhibullin}
\affil[1]{Spase Research Institute, Profsoyuznaya str. 84/32, 
Moscow  117997, Russia}
\affil[2]{Max-Planck-Institut f\"ur Astrophysik, Karl-
Schwarzshchild
  str. 1, Garching, Germany}
\affil[3]{INTEGRAL Science Data Center, 1290 Versoix, Switzerland}
\affil[4]{Kazan State University, 420008 Kazan, Kremlevskaya str. 
18,  Russia}
\affil[5]{Academy of Sciences of Tatarstan, 420000 Kazan, Bauman str. 20,  
Russia}

\begin{document}

\renewcommand{\topfraction}{1.0}
\renewcommand{\bottomfraction}{1.0}
\renewcommand{\textfraction}{0.0}


\maketitle

\begin{abstract}

We present the first results of the observations of the X-ray pulsars LMC
X-4, 4U0352+309 and EXO1722-363 performed with the INTEGRAL observatory. The
LMC X-4 was investigated during the whole superorbital cycle ($\sim$30 days)
and it was found that its period was not stable at this time scale. We
detected a variable X-ray flux (18-60 keV) from the pulsar EXO1722-363,
which could be connected with the orbital motion in the binary system.  A
more accurate position and the estimate of the orbital period for this
source are reported. We also investigated a hard X-ray spectrum of
4U0352+309 (X Persei) measured with INTEGRAL and report the detection of the
cyclotron absorption line at about 29 keV.

\end{abstract}

\section{Introduction}

At the moment more than 60 X-ray pulsars have been found to be members of
binary systems. Most of them are concentrated in the Galactic plane and
therefore are among the primary targets of the INTEGRAL observatory. We
investigated 15 X-ray pulsars --- persistent, transient, Be-systems etc.,
observed in different parts of the INTEGRAL observational programs.

Here we present preliminary results of the analysis of the INTEGRAL data for
several sources from this group.

\section{Observations and data analysis}

The international gamma-ray observatory INTEGRAL was launched by the Russian
launcher PROTON from the Baikonur cosmodrome on October 17, 2002 (Eismont et
al. 2003). The payload includes four principal instruments to carry out
simultaneous observations of sources in the X-ray, gamma-ray and optical
energy bands (Winkler et al. 2003). The observational program of the
observatory consists of Core and General Programs. The main part of Core
Program consist in regular scans of the galactic plane and deep observations
of the galactic central radian. General Program is formed from the proposals
of scientists. In this work we used the data collected by the detector ISGRI
(Lebrun et al. 2003) of the IBIS telescope (Ubertini et al. 2003) and JEM-X
telescope (Lund et al. 2003) in frames of both programs as well as during
calibration observations.

For imaging and spectral analysis of IBIS/ISGRI data we followed the methods
described in Revnivtsev et al (2004). For spectral analysis we used a simple
ratio of the flux measured from the source to the fluxes of the Crab nebula
in the same energy bands and assumed that the Crab nebula spectrum has the
form $d N(E)=10 E^{-2.1} dE$, where $N(E)$ is the number of photons at
energy $E$. Our analysis of an extensive set of Crab calibration
observations has shown that the source absolute flux can be recovered with
an accuracy of 10\% and the systematic uncertainty of relative flux
measurement in different energy channels is less than 5\%. Last value was
added to the following spectral analysis as a systematic uncertainty. The
analysis of the JEM-X data and the timing analysis of the ISGRI data were
performed with the standard package OSA 3.0.
\footnote{http://isdc.unige.ch/index.cgi?Soft+soft}.

\section{LMC X-4}

X-ray pulsar LMC X-4 ($\sim$13.5\,s) is a member of a massive binary system
with 1.408 day orbital period. In addition, LMC X-4 shows a modulation of
X-ray flux with 30.5 day cycle. During approximately 60\% of this cycle it
is observed in a high intensity state, while during the remaining 40\% of
the cycle its X-ray flux drops to $\sim 60$ times lower level (Lang et
al. 1981). This periodicity can be attributed to precessing tilted or warped
accretion disk, which periodically eclipses X-ray emitting regions.

\begin{figure}[]
  \centering
  \includegraphics[width=\columnwidth]{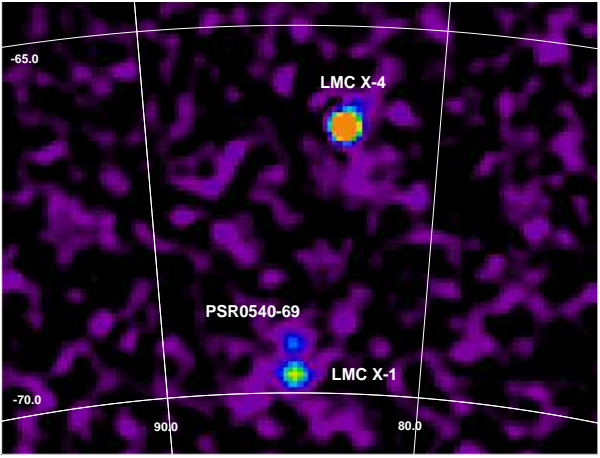}
  \caption{X-ray map of the region around of LMC X-4 obtained with 
    INTEGRAL/IBIS in the 18-60 keV energy band. \label{fig:lmc_ima}}
\end{figure}

LMC X-4 was observed with INTEGRAL in January 2003 during observations of the
Large Magellanic Cloud in the frame of the General Program. More than 1
million seconds of total useful exposure were collected. The X-ray map of the
region containing LMC X-4 obtained in the 18-60 keV energy band is shown in
Fig.\ref{fig:lmc_ima}.

\begin{figure}
  \centering
  \includegraphics[width=\columnwidth,bb=18 190 570 690 ]{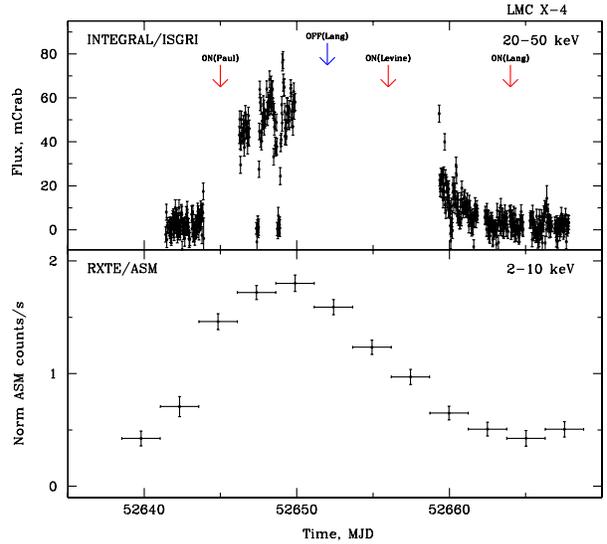}
  \caption{LMC X-4 30.5 d light curve, obtained with INTEGRAL/IBIS 
    in January 2003 (upper panel). The average for $\sim$8 years 30.5 day
    light curve, as measured with ASM/RXTE (bottom panel). The zero phase of
    the ASM/RXTE light curve was recalculated to date of INTEGRAL
    observations. \label{fig:lmc_30d}} 
\end{figure}

\begin{figure}[b]
  \centering
  \includegraphics[width=\columnwidth, bb=18 415 565 700]{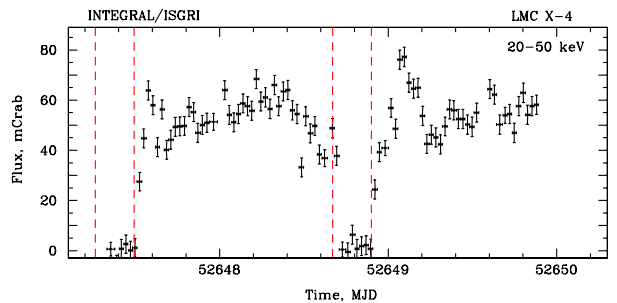}
  \caption{Orbital modulation of the hard X-ray flux of LMC X-4 in the high
    state. \label{fig:lmc_orb}} 
\end{figure}

These INTEGRAL observations covered almost whole 30.5 day cycle and allowed
us to investigate the source behavior in hard X-rays on different time
scales. In the upper panel of Fig.\ref{fig:lmc_30d} the light curve obtained
with IBIS/INTEGRAL in the 20-50 keV energy band is presented. The observed
turn-on moment does not coincide with those predicted from the previous
determinations of the 30.5 day cycle parameters (Lang et al.  1981, Levine
et al. 2000, Paul et al. 2002). The average epoch-folded light curve
obtained from about 8 years of ASM/RXTE observations is shown in the bottom
panel of Fig.\ref{fig:lmc_30d}. It is clearly seen that the average turn-on
moment does not coincide with the INTEGRAL one. Such discrepancy is the
evidence that the period of the 30.5 day cycle is not stable. It seems that
this instability of the precessing period is typical for all binary systems
where it is observed (see Clarkson et al. 2000). The orbital light curve of
LMC X-4 obtained with IBIS/INTEGRAL during two orbital cycles in the
high-state is shown in Fig.\ref{fig:lmc_orb}. Vertical dashed lines show the
moments of X-ray eclipses obtained from the known orbital parameters of the
system (Lang et al. 1981). It is interesting to note short increases of the
source intensity just after the eclipses, observed in both orbital cycles.

In the standard accretion torque model we can estimate the magnetic dipole
moment of LMC X-4 of the order of $\mu \sim 10^{31}$ G cm$^3$, which
corresponds to the magnetic field strength on the neutron star surface
$B\sim 10^{13}$\,G. If this is the case, cyclotron spectral feature is
expected at energies $E = 1.16 \cdot 10^{-11} B (1-{2GM\over{Rc2}})^{1/2}$
keV $\sim 100$ keV (here we take into account gravitational redshift of the
neutron star assuming its radius $R=12$\,km and mass $M=1.4M_\odot$).
Before INTEGRAL there were several discrepant indications of possible
cyclotron line feature in the spectrum of LMC X-4. Different authors (Levine
et al. 1991, Mihara 1995, Woo et al. 1996, La Barbera et al. 2001) claimed
the detection of a cyclotron absorption line in 19--100 keV energy range.

\begin{figure}[]
  \centering 
  \includegraphics[width=\columnwidth, bb=45 250 540
  700]{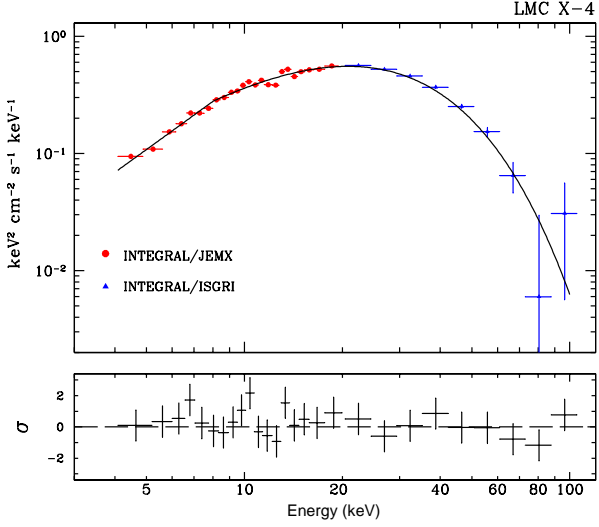}  
  \caption{Spectrum of LMC X-4 measured with INTEGRAL/JEM-X+IBIS (upper
    panel). Solid line represent a best-fit model described in the text.
    Deviations of the experimental data (in $\sigma$) from the best-fit
    model (bottom panel).  \label{fig:lmc_spec}}
\end{figure}

We performed analysis of the LMC X-4 broadband spectrum obtained with JEM-X
and IBIS telescopes of the INTEGRAL observatory during ~$\sim$~ 300 kseconds
of the source in the high state (Fig.\ref{fig:lmc_spec}). This analysis
shows that the LMC X-4 average spectrum can be well fitted by a
``canonical'' pulsar model (White et al. 1983), which consist of a simple
powerlaw with the photon index of $\Gamma=0.20\pm0.15$~ and a high energy
cutoff at $E_{cut}=9.1\pm0.8$~ keV with $E_{f}=11.0\pm0.6$~ keV
($\chi2=0.93$ per d.o.f.).

To search for a possible cyclotron resonance absorption line in the source
spectrum it was included into the model. A centroid energy of the line was
varied from 5 to 100 keV and other parameters were fitted. In this analysis
we have not found any indications of the cyclotron resonance absorption in
the source spectrum at the level more than $2\sigma$. Thus at the moment we
cannot confirm any results obtained earlier and make a final conclusion
about the magnetic field value in the source. More detailed analysis of the
INTEGRAL data is now in progress.

\section{EXO1722-363}

\begin{figure}
  \centering
  \includegraphics[width=\columnwidth]{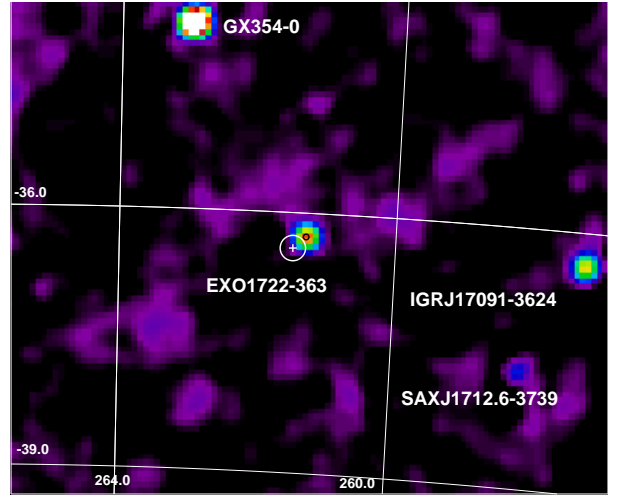}
  \caption{X-ray map of the EXO1722-363 region obtained with 
    INTEGRAL/IBIS in the 18--60\,keV energy band\label{fig:exo_ima}}
\end{figure}

The X-ray pulsar EXO1722-36 was regularly observed by INTEGRAL observatory
during the Core Program scans of the Galactic plane and during the ultra
deep observations of the Galactic Center. The source was clearly detected
with a high statistical significance at the energies up to $\sim$60
keV. Angular resolution, available with the INTEGRAL telescopes allowed us
to improve the localization of the source (see Lutovinov et al. 2003 for the
preliminary results) in comparison with that measured by the EXOSAT
observatory (Warwick et al. 1988). The X-ray map of sky around EXO1722-363
is shown in Fig.\ref{fig:exo_ima}. The white cross and the circle represent
the EXOSAT position and error circle. The new position measured with
IBIS/INTEGRAL, $RA=17^h25^m11^s$ and $Dec=-36^d16'30''$, is shown with the
black circle.

\begin{figure}[b]
  \centering
  \includegraphics[width=\columnwidth, bb=13 285 560 560]{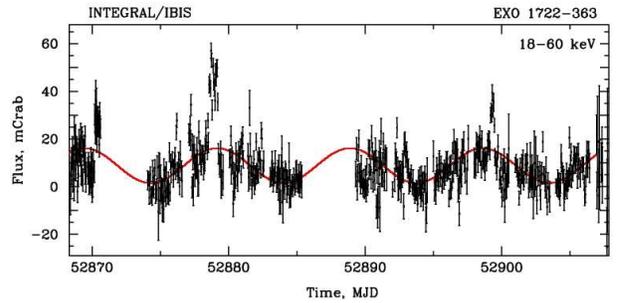}
  \caption{Light curve of EXO1722-363 in 18--60\,keV energy band. 
    Orbital modulations are clearly seen.\label{fig:exo_lc}}
\end{figure}

The long term observations of the source during the ultra deep exposure of
the Galactic Center revealed variability of the source X-ray flux on time
scales of several days. This variability observed with INTEGRAL/IBIS
(18--60\,keV) was also detected by RXTE in the 3--20\,keV energy band
(Markwardt, Swank 2003). The RXTE/PCA observations allowed to detect the
modulations of EXO1722-363 X-ray flux with a period of $9.737\pm0.004$ days
which was interpreted as an orbital period. Our measurements of the orbital
period from hard X-ray light curve (Fig.\ref{fig:exo_lc}) gives the value of
$9.725\pm0.045$ days, in good agreement with RXTE result.

\begin{figure}
  \centering
  \includegraphics[width=\columnwidth, bb=100 340 550 700]{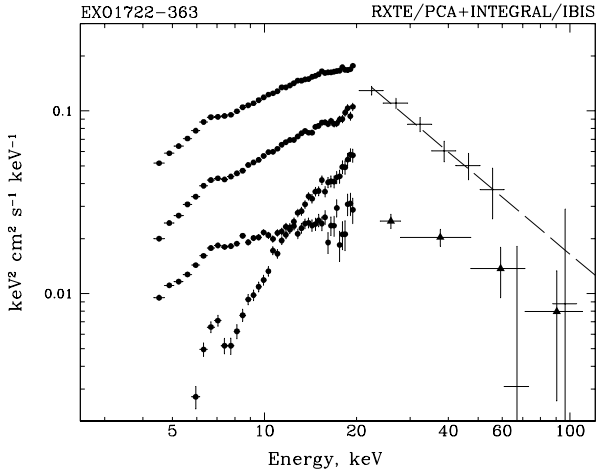}
  \caption{Broadband spectra of EXO1722-363, obtained by RXTE/PCA (dark
    circles), RXTE/HEXTE (triangles) and INTEGRAL/IBIS (crosses). Dashed
    line represent a best-fit approximation of the INTEGRAL/IBIS data by
    simple power law with the photon index of
    $\sim3.5$. \label{fig:exo_spec}}
\end{figure}

In order to construct the broadband energy spectrum of the source with
highest possible significance we combined the data of the INTEGRAL and RXTE
observatories. RXTE observatory performed a number of observations in
1998-2003 covering different orbital phases of the binary. The analysis of
the INTEGRAL/IBIS data showed that the shape of the hard X-ray part of the
spectrum did not change inspite of the strong flux variability (with
possible exception of the eclipse period), therefore in
Fig.\ref{fig:exo_spec} we present the hard X-ray part of the spectrum of EXO
1722-363 averaged over all available data (INTEGRAL/IBIS and
RXTE/HEXTE). However, low energy part of the spectrum (3--20 keV) strongly
depends on the orbital phase. The value of photoabsorption column strongly
varies with the orbital phase and can reach value of the order of
$N_H=10^{24}$ cm$^{-2}$ (see also Tawara et al. 1989). Examples of the
source spectrum during different orbital phases are presented in
Fig.\ref{fig:exo_spec}. It should be noted that the spectrum of the source
detected by the RXTE/PCA is strongly contaminated by a contribution of the
Galactic ridge emission. The spectra presented in Fig.\ref{fig:exo_spec}
were corrected for this contribution (using the method similar to that used
in Revnivtsev 2003). In this proceedings we will not go into the details of
the spectral analysis of the source.  A detailed analysis of the source
broadband spectral variability will be presented in a separate paper.

\section{4U0352+309/X Persei}
 
4U0352+309/X Persei is a low luminosity ($L_x\simeq$ 2 -- 4 $\times10^{34}$
erg/s) Be/X-ray binary, consisting of the 837 second X-ray pulsar 4U0352+309
and Be star companion optically identified as the star X Persei
(HD\,24534). Recent BeppoSAX observations showed that the spectrum of the
source can be described with a combination of the standard pulsar spectrum
model and a powerlaw with both high- and low-energy cut-offs, which
dominates above 15\,keV (di Salvo et al. 1998). The analysis of the RXTE
observations (Coburn et al. 2001) reveals the presence of the cyclotron line
at about 29\,keV. Be stars are known to have a dense slow disk-like
equatorial wind and a low-density fast wind from higher latitudes. The
bright state of the star correlates with the presence of the emission lines
and is the evidence of the presence of equatorial disk. No clear
correlations between the X-ray and the optical properties of the source were
found (e.g. Telting et al. 1998).
\begin{figure}[t]
  \centering
  \includegraphics[width=7cm,bb=85 265 500 710]{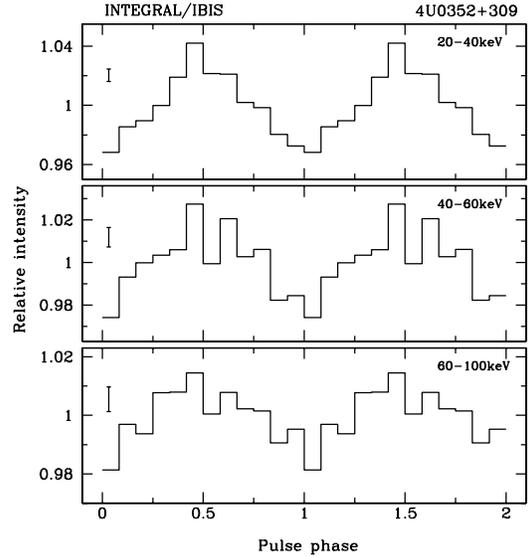}
  \caption {Pulse profile of 4U0352+309 in three energy bands, as seen with
    IBIS/INTEGRAL. Errors correspond to $1\sigma$ level. Background was not
    subtracted. 
    \label{fig:pers_pp}}
\end{figure}

\begin{figure}b]
  \centering
  \includegraphics[width=7cm, bb=50 50 560 580]{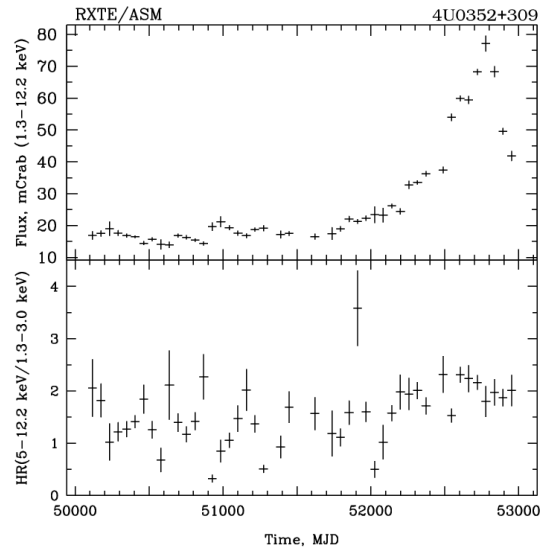}
  \caption{4U0352+309 light curve in the 2-12 keV energy band as seen 
    by ASM/RXTE during $\sim$8 years (upper panel) with corresponding 
    hardness ratio (bottom panel). 
    \label{fig:pers_asm}}
\end{figure}

4U0352+309 was in the field of view of the INTEGRAL/IBIS telescope on August
14, 2003 during the calibration observations of the Crab nebula. The total
useful exposure time was about 50 kseconds.The source was detected with an
average flux of $\sim40$\,mCrab in the 20--100\,keV energy band. The
coherent pulsations with a period of $\sim838$\,s were detected at the
energies up to 100\,keV (Fig.\ref{fig:pers_pp}). The pulse profile has a
typical shape for this source with one broad peak. Unfortunately, we can
only very roughly estimate the source pulse fraction in these energies as
$\sim20$\% because of a lack of accurate knowledge of the background.
\begin{figure}[t]
  \centering
  \includegraphics[width=7cm, bb=50 190 515 695]{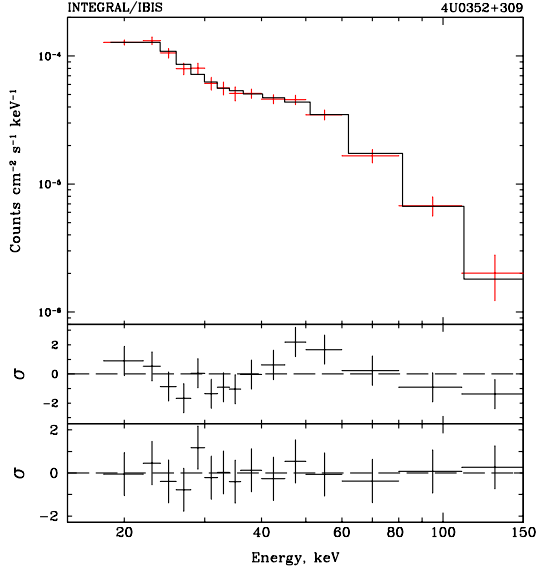}
  \caption{Spectrum of 4U0352+309 obtained with IBIS/INTEGRAL (upper 
    panel). Histogram represents a best-fit model of the power law with the
    high-energy cutoff and the cyclotron line. Middle and bottom panels show
    the deviations of the experimental data (in $\sigma$) from best fits
    for the two model spectra: without cyclotron absorption line and with
    this line, respectively.
    \label{fig:pers_spc}}
\end{figure}

\begin{figure}[b]
  \centering
  \includegraphics[width=7cm, bb=45 275 515 695]{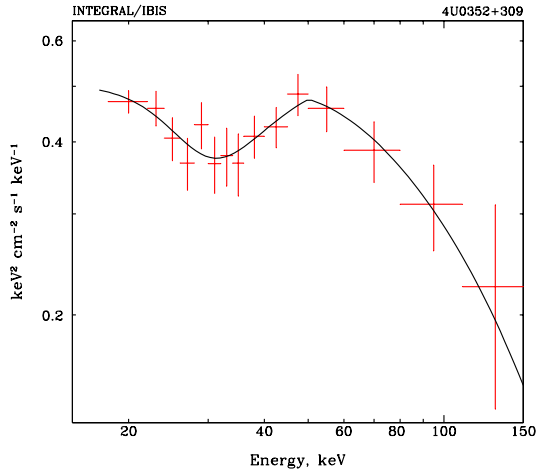}
  \caption{Energy spectrum of 4U0352+309. Histogram represents best-fit 
    model of power law with the high-energy cutoff and absorption cyclotron line.
    \label{fig:pers_eeufs}}
\end{figure}

Analysis of the 4U0352+309 light curve obtained with the ASM/RXTE monitor
indicates that during last 2 years source intensity in 1.3--12\,keV energy
band increased by several times and reached its maximum on May 2003 without
strong changes in the hardness (Fig.\ref{fig:pers_asm}). Thus observations
of 4U0352+309 with INTEGRAL were performed closely to the maximum of the
source intensity.

The source spectrum measured with IBIS/INTEGRAL is shown in the upper panel
of Fig.\ref{fig:pers_spc}. The dots with error bars represent the
pulse-height spectrum (in units of cnts\,s$^{-1}$\,cm$^{-2}$\,keV$^{-1}$),
and the histogram represents the corresponding model spectrum of the
powerlaw with a high energy cutoff which is typical for X-ray pulsars (White
et al. 1983). The deviations of the measured spectrum from the model spectra
convolved with the instrumental response matrix are shown in the middle
panel. It is clearly seen that near 30\,keV there is a wide absorption
feature, which can be interpreted as a cyclotron resonance absorption
line. Addition of this component to the model improves the fit (bottom
panel) with a significance of $\sim3\sigma$ ($\Delta \chi2$-test). The
best-fit parameters are: photon index $\Gamma=1.92\pm0.19$,
$E_{cut}=50\pm16$ keV, $E_{f}=77\pm27$ keV, $\tau=0.33\pm0.12$,
$E_{cycl}=28.8\pm2.5$ keV (the cyclotron line width was taken from Coburn et
al.(2001) and fixed on 9 keV).  The obtained centroid energy of the
cyclotron line is in a well agreement with RXTE results (Coburn et
al. 2001). The energy spectrum of 4U0352+309 is shown in
Fig.\ref{fig:pers_eeufs}.

\begin{figure}[t]
  \centering
  \hspace{0mm}\includegraphics[width=\columnwidth, bb=45 350 550
  690]{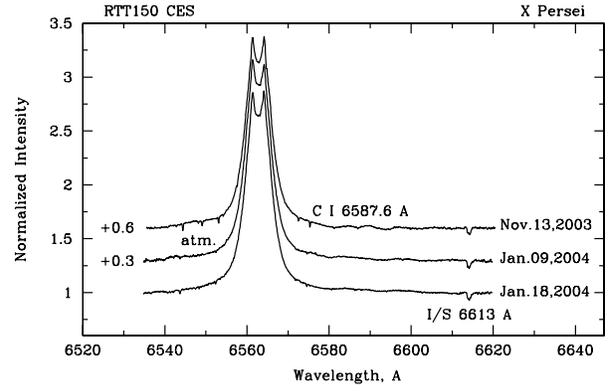}  
  \caption{The double peaked H$\alpha$ line profiles measured with  
    RTT-150 from X Persei.   
    \label{fig:pers_ha}} 
\end{figure}

We have obtained series of high-resolution optical spectra of X Persei with
1.5-m Russian-Turkish Telescope (RTT150), using its new Coude-echelle high
resolution spectrometer. Three high signal-to-noise spectra have been
obtained in November 13/14, 2003, January 9/10, 2004 and January 18/19, 2004
in the spectral range 4000-9000 A with a resolution of $R = 40000$. Profiles
of $H_\alpha$-line measured in these observations are shown in
Fig.\ref{fig:pers_ha}. The registered intensity of $H_\alpha$ line emission
was very high and it indicates that the optical star of X Persei binary
system was in an active state during the observations.

\begin{figure}[] 
  \centering 
  \hspace{0mm}\includegraphics[width=\columnwidth, bb=45 300 550
  690]{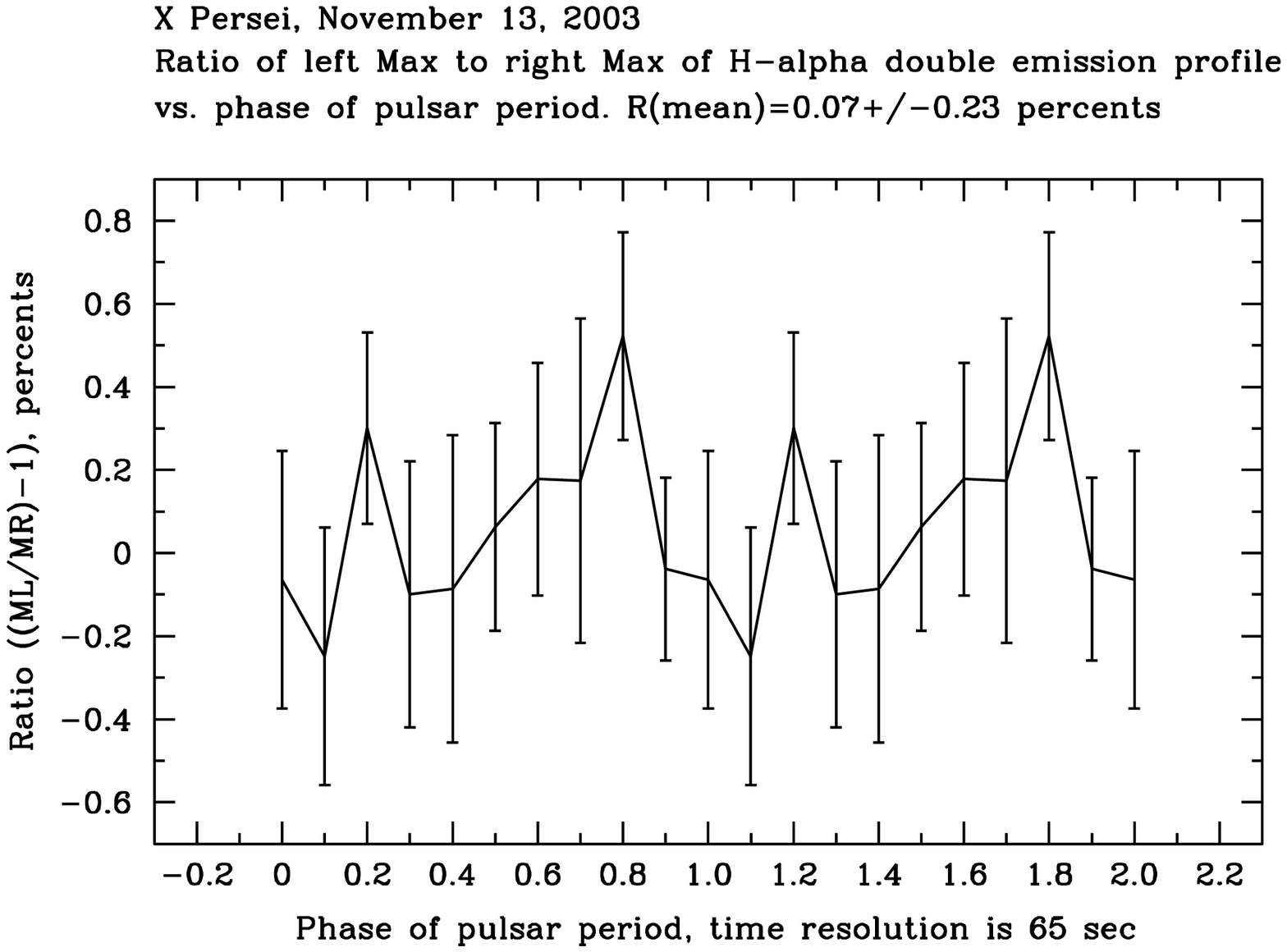} 
  \caption{Dependence of the ratio of the intensities of the blue  
    and red-shifted peaks of the H$\alpha$ line on the pulse period phase.
    \label{fig:pers_har}} 
\end{figure} 

To search for a possible modulation of the $H\alpha$-line intensity on a
time scale of the pulsar period a set of 200 short (30 seconds) exposure
spectra was obtained on November 13/14, 2003 with the resolution of $R =
20000$. We selected for the following analysis 107 spectra with highest S/N
ratio. For each spectrum we calculated a ratio of left-to-right peaks
intensities of the $H_\alpha$-line profile. The obtained series cover
approximately 8 pulsar periods. All data were binned into 10 pulse phase
bins and averaged inside of each of them. The ratio of left-to-right peaks
intensities of the $H_\alpha$-line profile versus a pulse period phase is
presented in Fig.\ref{fig:pers_har}. No significant modulation of the
measured value with the pulse period was found.

\section{Summary}

>From our preliminary analysis of X-ray pulsars, observed with the INTEGRAL
observatory we can conclude:

\begin{itemize}

\item no cyclotron line was found in the LMC X-4 spectrum in the 5-100 keV
  energy band;
\item new accurate position was measured for EXO1722-363; possible orbital
  period $\sim9.73$ day was found; 
\item unusually bright state of 4U0352+309/X-Persei system was observed;
  cyclotron resonance absorption feature is detected at $\sim28.8$ keV; no
  pulsations of $H_\alpha$ emission line flux were detected.
\end{itemize}

\section*{Acknowledgements}

Authors thank E.Churazov for the developing of the methods of the analysis
of the IBIS data and software. We are thankful to the referee, Andrea
Santangelo, for comments which helped to improve the paper. Research has
made use of data obtained through the INTEGRAL Science Data Center (ISDC),
Versoix, Switzerland, Russian INTEGRAL Science Data Center (RSDC), Moscow,
Russia, and High Energy Astrophysics Science Archive Research Center Online
Service, provided by the NASA/Goddard Space Flight Center. Work was
partially supported by grants of Minpromnauka NSH-2083.2003.2,
NSH-1789.2003.2 and 40.022.1.1.1102, RFBR 04-02-17276 and program of Russian
Academy of Sciences ``Non-stationary phenomena in astronomy''. This work was
partially made during visits to the International Space Science Institute
(ISSI), Bern which AL, MR, MC, IB, SM, RB, MN thank for the hospitality and
financial support. RTT150 Coude-echelle spectrometer has been created under
the support of Academy of Science of Tatarstan (AST). IB and NS are grateful
to AST for this financial support.

\end{document}